\documentclass[twocolumn]{aastex631}

\shorttitle{The orbit of WASP-106\,b is aligned with its star}
\shortauthors{Jan-Vincent Harre et al.}

\begin{document} 

\title{The Orbit of Warm Jupiter WASP-106\,b is aligned with its Star}
   %\title{The Projected Obliquity of the Warm Jupiter WASP-106\,b}

   %\subtitle{I. Overviewing the $\kappa$-mechanism}

\author[0000-0001-8935-2472]{Jan-Vincent Harre}
\affiliation{Institute of Planetary Research, German Aerospace Center (DLR), Rutherfordstraße 2, 12489 Berlin, Germany}

\author[0000-0002-2386-4341]{Alexis M. S. Smith}
\affiliation{Institute of Planetary Research, German Aerospace Center (DLR), Rutherfordstraße 2, 12489 Berlin, Germany}

\author[0000-0003-3618-7535]{Teruyuki Hirano}
\affiliation{Astrobiology Center, 2-21-1 Osawa, Mitaka, Tokyo 181–8588, Japan}
\affiliation{National Astronomical Observatory of Japan, 2-21-1 Osawa, Mitaka, Tokyo 181–8588, Japan}
\affiliation{Department of Astronomical Science, School of Physical Sciences, The Graduate University for Advanced Studies (SOKENDAI), 2-21-1, Osawa, Mitaka, Tokyo, 181–8588, Japan}

\author[0000-0001-6803-9698]{Szilárd Csizmadia}
\affiliation{Institute of Planetary Research, German Aerospace Center (DLR), Rutherfordstraße 2, 12489 Berlin, Germany}

\author[0000-0002-5510-8751]{Amaury H. M. J. Triaud}
\affiliation{School of Physics \& Astronomy, University of Birmingham, Edgbaston, Birmingham B15 2TT, United Kingdom}

\author[0000-0001-7416-7522]{David R. Anderson}
\affiliation{Centre for Exoplanets and Habitability, University of Warwick, Coventry CV4 7AL, UK}
\affiliation{Department of Physics, University of Warwick, Coventry CV4 7AL, UK}

%   \author{J.-V. Harre\inst{1}\thanks{E-mail: jan-vincent.harre@dlr.de} 
%   $^{\href{https://orcid.org/0000-0001-8935-2472}{\includegraphics[scale=0.5]{figures/orcid.jpg}}}$ \and 
%   T. Hirano\inst{2} \and
%A. M. S. Smith\inst{1} $^{\href{https://orcid.org/0000-0002-2386-4341}{\includegraphics[scale=0.5]{figures/orcid.jpg}}}$ \and
%Sz. Csizmadia\inst{1} \and
%A. H. M. J. Triaud \and
%D. R. Anderson
%}

%\inst{\label{inst:1} Institute of Planetary Research, German Aerospace Center (DLR), Rutherfordstrasse 2, 12489 Berlin, Germany \and
%\label{inst:2} Astrobiology Center, NINS, Mitaka, Tokyo, Japan (UPDATE PLS!)}

%\authorrunning{J.-V. Harre et al.}
%\titlerunning{The orbit of WASP-106\,b is aligned}
%   \date{Received DD MM, 2023; accepted DD MM, 2023}

% \abstract{}{}{}{}{} 
% 5 {} token are mandatory
 
 \begin{abstract}
     Understanding orbital obliquities, or the misalignment angles between a star's rotation axis and the orbital axis of its planets, is crucial for unraveling the mechanisms of planetary formation and migration.
     In this study, we present an analysis of Rossiter-McLaughlin (RM) observations of the warm Jupiter exoplanet WASP-106\,b. The high-precision radial velocity measurements were made with HARPS and HARPS-N during the transit of this planet. We aim to constrain the orientation of the planet's orbit relative to its host star's rotation axis.
     The RM observations are analyzed using a code which models the RM anomaly together with the Keplerian orbit given several parameters in combination with a Markov chain Monte Carlo implementation.
     We measure the projected stellar obliquity in the WASP-106 system for the first time and find $\lambda = (-1\pm11)^\circ$, supporting the theory of quiescent migration through the disk.
 \end{abstract}
   
   \keywords{Exoplanet evolution --- Exoplanet migration --- Radial velocity --- Photometry --- Transits}

%
%-------------------------------------------------------------------

\section{Introduction}
% Obliquity measurements of warm Jupiters are ...
% important because we can test migration mechanisms in simualtions our test ouir theories in the data
% Are WJ and HJ formed differently, are they distinct populations?
% What could be mechanisms that lead to HJ or WJ
% Maybe mention the existence of other planets in those systems, like W43 for HJs as a special one, WJ often more planets?
% only a small sample of WJs with measured obliquities

% Cite Albrecht review paper and Rice paper
% Mention stellar multiplicity and that for single stars they look mostly aligned from that paper
% Maybe mention the 90° tendency for warm Jupiters from the Rice? paper
% Mention new paper with misaligned WJ, and also mention the difference between the projected and 3D obliquity.

The diversity of exoplanetary systems has unveiled a wealth of information about the complex interplay between planets and their host stars. One key aspect that has captured considerable attention in recent years is the obliquity of planetary orbits, which plays a major role in understanding the mechanisms governing planetary formation, migration, and long-term stability. The projected obliquity, defined as the misalignment angle between the stellar rotation axis and the planetary orbital axis, offers valuable insights into the dynamical history of these systems.

In recent years, studies focussing on hot and warm Jupiter systems have revealed a wide range of obliquity measurements, with some planets exhibiting near perfect alignment with their host star's equator (e.g. \citealt{2013ApJ...772...80F, 2015ApJ...800L...9A, 2017A&A...601A..53E, 2017AJ....154..194L, 2021AJ....162..256W}), while others display significant misalignments, polar or even retrograde orbits (e.g. \citealt{2010ApJ...722L.224B, 2011A&A...527L..11H, 2012ApJ...744..189A, 2017AJ....153...94C, 2020AJ....160....4A}). These observations have led to various theories regarding the formation and subsequent migration processes of these gas giants. However, hot Jupiters are seemingly misplaced by high-eccentricity migration, with spin-orbit misalignments being able to be reverted by tidal damping \citep{2022ApJ...926L..17R} if they are orbiting cool stars. For hot hosts ($T_\mathrm{eff} > 6250\,K$) of hot Jupiters it is suggested that due to their thinner convective zones they are not able to realign misaligned orbits efficiently \citep{2010ApJ...718L.145W}.

Warm Jupiters, especially those in single star systems, orbit preferentially in alignment with their host star's rotation \citep{2022AJ....164..104R}, even though there has recently been a discovery of a warm Jupiter with a significant misalignment of $\lambda = 38.9^{+2.8\,\circ}_{-2.7}$ and eccentricity $(e=0.57^{+0.12}_{-0.16})$ \citep{2023arXiv230516495D}. Understanding the origins of such diverse obliquity distributions and their implications for planetary migration mechanisms remain active areas of research. For a recent review on stellar obliquities see e.g. \citep{2022PASP..134h2001A}.

%\JH{Mention difference of projected vs 3D obliquities!}.
Lately, there have been numerous measurements of the true 3D obliquities which require the stellar inclination to be measured in addition to the projected obliquity, see e.g. \cite{2016A&A...588A.127C, 2023A&A...669A..63B, 2023MNRAS.522.4499D}. Depending on the orientation of the stellar spin axis with regard to our line of sight to the observed system, the projected and 3D obliquity can be quite different from one another (see e.g. \cite{2023ApJ...949L..35H}).

In this paper, we present an investigation into the projected obliquity of one warm Jupiter, namely WASP-106\,b, utilising a combination of high-precision radial velocity (RV) measurements and the Rossiter-McLaughlin (RM) effect \citep{1924ApJ....60...15R, 1924ApJ....60...22M}. By analysing the stellar radial velocity variations during planetary transit, we aim to derive a precise measurement of the projected obliquity for this system. 
This is particularly interesting for the examined system, since the host star's effective temperature is just at the Kraft break \citep{1967ApJ...150..551K}, which separates stars with deep and shallow convective zones.
These measurements offer valuable constraints on the orientations of the planetary orbits and provide insights into their formation histories, migration pathways, and subsequent dynamical evolution.

Our study contributes to the growing body of knowledge surrounding obliquity measurements and their significance in the context of planetary systems. By focussing on warm Jupiter exoplanets, we aim to deepen our understanding of the mechanisms that govern the formation and migration of these gas giants in close proximity to their host stars. Through this research, we want to find out how these systems acquired their observed obliquities and the role played by various dynamical processes in shaping their architectures.

Overall, this work represents a step forward in characterizing the obliquity distributions of warm Jupiter systems, by expanding the limited sample of about 16 measured obliquities of warm Jupiters (see e.g. \citealt{2022AJ....164..104R, 2022A&A...664A.162M, 2023arXiv230516495D, 2023arXiv230707598S}), and provides valuable insights into the complex interplay between planetary migration, dynamical interactions, and the properties of host stars.

In Sect.~\ref{sec:obs} we describe the set of observations that are used in our analysis. Sect.~\ref{sec:methods} gives information about our RM modelling code, with the results presented in Sect.~\ref{sec:results} and discussed in Sect.~\ref{sec:disc}. The conclusions are given in Sect.~\ref{sec:conclusion}.

%--------------------------------------------------------------------

\section{Observations\label{sec:obs}}

\subsection{System parameters}
The stellar host of WASP-106\,b is an F9-type star.
%with an effective temperature of $6265\pm36$\,K, a mass of $(1.26\pm0.05)$\,M$_\odot$ and a radius of $(1.42\pm0.03)$\,R$_\odot$. 
The warm Jupiter in this system is on a $9.29$\,d orbit, has a mass of two times that of Jupiter,
%and radius of $(2.00\pm0.08)$\,M$_\mathrm{J}$ and $(1.10\pm0.02)$\,R$_\mathrm{J}$, respectively \JH{Do I even put all this here since it's also in the table?}, 
and was discovered by \citet{2014AA...570A..64S}. A summary of the stellar parameters can be found in Table~\ref{tab:star_table}. \\

% \begin{deluxetable}{lcc}
% %\tabletypesize{\scriptsize}
% \tablewidth{0pt}
% \tablecaption{Summary of stellar properties of WASP-106.\label{tab:star_table}}
% \tablehead{
% \colhead{Parameter} & \colhead{Value} & \colhead{Source}
% }
% \startdata
%     RA (J2000) & 11:05:43.14 & Simbad \\
%     DEC (J2000) & -05:04:45.94 & Simbad \\
%     $\mu_\mathrm{RA}$ [mas\,yr$^{-1}$] & -24.73 & Gaia DR3 \\
%     $\mu_\mathrm{DEC}$ [mas\,yr$^{-1}$] & -13.13 & Gaia DR3 \\
%     Age [Gyr] & $7\pm2$ & \citep{2021MNRAS.506.3810B} \\
%     Parallax [mas] & 2.783$\pm$0.018 & Gaia DR3 \\
%     Distance [pc] & 351.8$^{+2.0}_{-2.2}$ & Gaia DR3 \\
%     V [mag] & 11.21 & Simbad \\
%     G [mag] & 11.36 & Gaia DR3 \\
%     Spectral type & F9 D & Simbad \\
%     $T_\mathrm{eff}$ [K] & $6265\pm36$ & \citep{2021MNRAS.506.3810B} \\
%     $v\sin i$ [km$\,$s$^{-1}$] & $6.3\pm0.7$ & \citep{2014AA...570A..64S} \\
%     $\log\,g$ & $4.38^{+0.04}_{-0.04}$ & \citep{2021MNRAS.506.3810B} \\
%     $\rho_\star$ [$\rho_\odot$] & $0.81\pm0.15$ & \citep{2021MNRAS.506.3810B} \\
%     $\left[\mathrm{Fe/H}\right]$ [dex] & $0.15\pm0.03$ & \citep{2021MNRAS.506.3810B} \\
%     $R_\star$ [$R_\odot$] & $1.42\pm0.02$ & \citep{2021MNRAS.506.3810B} \\
%     $M_\star$ [$M_\odot$] & $1.26\pm0.05$ & \citep{2021MNRAS.506.3810B} \\
% \enddata
% \end{deluxetable}

\begin{table}[h]
    \centering
    \caption{Summary of stellar properties of WASP-106.}
    \begin{tabular}{l c c}
	\hline
	Parameter & Value & Source \\\hline
    RA (J2000) & 11:05:43.14 & Simbad \\
    DEC (J2000) & -05:04:45.94 & Simbad \\
    $\mu_\mathrm{RA}$ [mas\,yr$^{-1}$] & -24.73 & Gaia DR3 \\
    $\mu_\mathrm{DEC}$ [mas\,yr$^{-1}$] & -13.13 & Gaia DR3 \\
    Age [Gyr] & $7\pm2$ & \citep{2021MNRAS.506.3810B} \\
    Parallax [mas] & 2.783$\pm$0.018 & Gaia DR3 \\
    Distance [pc] & 351.8$^{+2.0}_{-2.2}$ & Gaia DR3 \\
    V [mag] & 11.21 & Simbad \\
    G [mag] & 11.36 & Gaia DR3 \\
    Spectral type & F9 D & Simbad \\
    $T_\mathrm{eff}$ [K] & $6265\pm36$ & \citep{2021MNRAS.506.3810B} \\
    $v\sin i$ [km$\,$s$^{-1}$] & $6.3\pm0.7$ & \citep{2014AA...570A..64S} \\
    log\,$g$ & $4.38^{+0.04}_{-0.04}$ & \citep{2021MNRAS.506.3810B} \\
    $\rho_\star$ [$\rho_\odot$] & $0.81\pm0.15$ & \citep{2021MNRAS.506.3810B} \\
    $\left[\mathrm{Fe/H}\right]$ [dex] & $0.15\pm0.03$ & \citep{2021MNRAS.506.3810B} \\
    $R_\star$ [$R_\odot$] & $1.42\pm0.02$ & \citep{2021MNRAS.506.3810B} \\
    $M_\star$ [$M_\odot$] & $1.26\pm0.05$ & \citep{2021MNRAS.506.3810B} \\\hline
   \end{tabular}
   \label{tab:star_table}
\end{table}

\subsection{Radial velocity data}
We obtained archival HARPS\footnote{Freely available at the ESO archive \href{http://archive.eso.org/eso/eso_archive_main.html}{http://archive.eso.org/}.} \citep{2003Msngr.114...20M} and HARPS-N\footnote{Freely available at the TNG archive \href{http://archives.ia2.inaf.it/tng/}{http://archives.ia2.inaf.it/}.} \citep{2012SPIE.8446E..1VC} Rossiter-McLaughlin observations, which are described in Table~\ref{tab:RM_obs}, for our target, as well as further publicly available out-of-transit data from the literature.
In addition to the RM observations, for WASP-106\,b, there are 29 RVs available in \citet{2014AA...570A..64S}. Of these, 20 observations were obtained using the CORALIE spectrograph \citep{2000A&A...354...99Q} and 9 using the SOPHIE spectrograph \citep{2008SPIE.7014E..0JP}, all of which were taken between January 2013 and February 2014. The radial velocity data can be found in Table~\ref{tab:RVs}. \\

%WASP-106:
%RM 2014-04-03 until 04 - 21x HARPS; Prog.ID. 093.C-0474(A)\\
%RM 2015-01-16 until 17 - 33x HARPS; Prog.ID. 094.C-0090(A)\\
%RM 2015-02-13 until 14 - 48x HARPS-N; Prog.ID. OPT14B\_66\\
%Also public data from previous publications.
%Additionally:
%(Observed with ESPRESSO: 38x on 2023-03-18 (night of 17th), Prog.ID. 110.23Y8.002).\\

% \begin{deluxetable}{cccc}
% %\tabletypesize{\scriptsize}
% \tablewidth{0pt}
% \tablecaption{Observation log for Rossiter-McLaughlin observations of our target WASP-106\,b.\label{tab:RM_obs}}
% \tablehead{
% \colhead{Instrument} & \colhead{Night} & \colhead{Exposures} & \colhead{Programme ID}
% }
% \startdata
%         HARPS & 2014-04-03 & 21 & 093.C-0474(A)\\
%         HARPS & 2015-01-16 & 33 & 094.C-0090(A)\\
%         HARPS-N & 2015-02-13 & 48 & OPT14B\_66\\
% \enddata
% %\vspace{-0.5cm}
% \end{deluxetable}

\begin{table*}[]
    \centering
    \renewcommand{\arraystretch}{1.1} % Default value: 1
    \setlength{\tabcolsep}{12pt}
    \caption{Observation log for Rossiter-McLaughlin observations for our target WASP-106\,b.}
    \begin{tabular}{c c c c c}
        \hline
        Instrument & Night & Exposures & Programme ID \\\hline
        HARPS & 2014-04-03 & 21 & 093.C-0474(A)\\
        HARPS & 2015-01-16 & 33 & 094.C-0090(A)\\
        HARPS-N & 2015-02-13 & 48 & OPT14B\_66\\[4pt] \hline
    \end{tabular}
    %\tablecomments{These data are freely available at the TNG archive\footnote{\href{http://archives.ia2.inaf.it/tng/}{http://archives.ia2.inaf.it/}} in the case of HARPS-N and at the ESO archive\footnote{\href{http://archive.eso.org/eso/eso_archive_main.html}{http://archive.eso.org/}} for HARPS observations.}
    \tablecomments{These data are freely available at the \href{http://archives.ia2.inaf.it/tng/}{TNG archive} in the case of HARPS-N and at the \href{http://archive.eso.org/eso/eso_archive_main.html}{ESO archive} for HARPS observations.}
    \label{tab:RM_obs}
\end{table*}

\subsection{Photometric data}
In addition, the investigated system has been observed by TESS (Transiting Exoplanet Survey Satellite). These data are publicly available at the MAST Portal under \dataset[10.17909/wa2p-c522]{http://dx.doi.org/10.17909/wa2p-c522}.
%\footnote{\href{https://mast.stsci.edu/portal/Mashup/Clients/Mast/Portal.html}{https://mast.stsci.edu/}}.
WASP-106 was observed in Sectors 9, 36, 45 and 46. The cadences are 120\,s each, except for the Sector 36 observations, where data with a cadence of 20\,s are also available.

%------------------------------------------------------------------

\section{Modelling\label{sec:methods}}
\subsection{Radial velocities}
% From Teru:
To fit the observed RVs during and out of transit, we used a C++ script, which 
implements the analytic calculation for the RM effect by \citet{2011ApJ...742...69H} in addition to the standard RV curve from the Keplerian orbit of the planet. 
Based on the assumption that the RVs are derived by the template-matching 
technique as for e.g. the HARPS DRS (Data Reduction Software) or SERVAL \citep{2018A&A...609A..12Z}, the script returns the 
velocity anomaly due to the RM effect and the Keplerian orbit for a set of times sampled. The input 
parameters to the script are all RV, RM and transit parameters, namely the RV amplitude $K$, the eccentricity $e$ and argument of periastron $\omega$ as $\sqrt{e}\,\sin\left(\omega\right)$ and $\sqrt{e}\,\cos\left(\omega\right)$, the stellar rotational velocity $v\,\sin\,i$, the projected stellar obliquity angle $\lambda$, the quadratic limb-darkening parameters $u_1$ and $u_2$, the RV offset $\gamma$, the transit center time $T_\mathrm{c}$, the scaled semi-major axis $a\,R_\star^{-1}$, impact parameter $b$, planet-to-star radius ratio $R_\mathrm{p}\,R_\star^{-1}$, planetary orbital period $P$, the number of discrete times to compute the model at, and lastly, the discrete times to evaluate the model at.
% \JH{describe which ones?} plus two RM parameters, 
% i.e., the projected stellar obliquity $\lambda$ and rotation velocity $v\sin i$. 
The macroturbulent velocity $\zeta$ needed to compute the analytic RM 
model is set to $3.8$ km s$^{-1}$ \citep{1977ApJ...218..530G, 2018ApJ...857..139G} as WASP-106 is a solar-type star. 
The RM model from the modelling code is then optimised to fit to our data using an MCMC algorithm, implemented using the \textit{emcee} \textit{Python} package \citep{2013PASP..125..306F}. We use 100 walkers with 25.000 steps and a burn-in period of 5000 steps. To ensure convergence, we compute the autocorrelation time and compare it to the number of steps divided by 50.
In our MCMC optimization, we used Gaussian priors from the discovery and follow-up paper, except for the mid-transit time $T_\mathrm{c}$ and orbital period $P$, which we infer from the TESS photometry, and the projected stellar obliquity $\lambda$ and the specific RV offsets $\gamma_i$ where we apply uniform priors, allowing all possible values, see Table~\ref{tab:priors_results}.
The limb-darkening parameters were taken from \citet{2013A&A...552A..16C} with the closest pass band in the case of the HARPS and HARPS-North spectrographs being the V- and y-bands, both of which are almost identical to each other for our stellar parameters.

\begin{table*}[]
    \centering
    \renewcommand{\arraystretch}{1.1} % Default value: 1
    \setlength{\tabcolsep}{16pt}
    \caption{Priors for and results from our MCMC modelling of the RV and the photometric TESS data. In the case of a normal (Gaussian) distribution ($\mathcal{N}$), the values within the brackets refer to the mean and standard deviation, respectively. For the uniform distribution ($\mathcal{U}$), the values refer to the lower and upper boundaries. The prior values are derived from those stated in \citet{2014AA...570A..64S} and \citet{2021MNRAS.506.3810B} for WASP-106\,b, except for $T_\mathrm{c}$ and $P$, which were determined in this paper using the TESS photometry. The quadratic limb darkening parameters were derived from the V- and y-band theoretical values from \citet{2013A&A...552A..16C}, which are a close match to the HARPS/HARPS-N wavelength range. $T_\mathrm{c}$ is given in BJD$_\mathrm{TDB}$ - 2450000. The subscripts ``H1'', ``H2'', ``HN'', ``CO'' and ``SO'' stand for HARPS first observation, HARPS second observation, HARPS-N, CORALIE and SOPHIE, respectively.}
    \begin{tabular}{l c c c}
        \hline
        Parameter [unit] & Prior & RV Result & TESS Phot. Result\\\hline
        $K$ [m\,s$^{-1}$] & $\mathcal{N}\,(165.3, 4.3)$ &  $162.5\pm1.5$ & -\\
        $\sqrt{e}\cos\omega$ & $\mathcal{N}\,(0.0, 0.3)$ & $-0.171\pm0.015$ & -\\
        $\sqrt{e}\sin\omega$ & $\mathcal{N}\,(0.0, 0.3)$ & $0.152\pm0.029$ & -\\
        $v\sin i$ [m$\,$s$^{-1}$] & $\mathcal{N}\,(6300,700)$ & $6714.1\pm251.8$ & -\\
        $\lambda$ [$^\circ$]& $\mathcal{U}\,(-180,180)$ & $-1.05\pm11.31$ & -\\
        $u_1$ & $\mathcal{N}\,(0.43, 0.15)$ & $0.787\pm0.109$ & $0.214\pm0.186$\\
        $u_2$ & $\mathcal{N}\,(0.24, 0.15)$ & $ 0.400\pm0.138$ & $0.070\pm0.215$\\
        $a\,R_\star^{-1}$ & $\mathcal{N}\,(14.2, 0.4)$ & $14.04\pm0.28$ & $13.9\pm0.07$\\
        $b$ & $\mathcal{N}\,(0.13,0.17)$ & $0.036\pm0.092$ & $0.274\pm0.047$\\
        $R_\mathrm{p}\,R_\star^{-1}$ & $\mathcal{N}\,(0.078, 0.001)$ & $0.0780\pm0.0010$ & $0.0760\pm0.0017$\\
        $T_\mathrm{c}$ [d] & $\mathcal{N}\,(9297.11711, 0.00077)$ & $9297.11710\pm0.00076$ & $9297.11711\pm0.00077$\\
        $P$ [d] & $\mathcal{N}\,(9.289711, 0.000019)$ & $9.289718\pm0.000005$ & $9.289711\pm0.000019$\\ 
        $\gamma_\mathrm{H1}$ [km\,s$^{-1}$] & $\mathcal{U}\,(17.0,17.5)$ & $17260.1\pm1.2$ & -\\
        $\gamma_\mathrm{H2}$ [km\,s$^{-1}$] & $\mathcal{U}\,(17.0,17.5)$ & $17237.5\pm0.9$ & -\\
        $\gamma_\mathrm{HN}$ [km\,s$^{-1}$] & $\mathcal{U}\,(17.0,17.5)$ & $ 17259.9 \pm 1.2$ & -\\
        $\gamma_\mathrm{CO}$ [km\,s$^{-1}$] & $\mathcal{U}\,(17.0,17.5)$ & $17248.1 \pm 4.3$ & -\\
        $\gamma_\mathrm{SO}$ [km\,s$^{-1}$] & $\mathcal{U}\,(17.0,17.5)$ & $17190.7 \pm 5.7$ & -\\
         \hline
    \end{tabular}
    \label{tab:priors_results}
\end{table*}

\subsection{TESS photometry}

We model the TESS photometry using the Transit and Light Curve Modeller, version 12, (TLCM, \citealt{2020MNRAS.496.4442C}). TLCM is a modelling code that allows analysis, fitting and simulation of light curves of transiting exoplanets, as well as radial velocities. The code employs the \citet{2002ApJ...580L.171M} descriptions of transits and occultations, with numerous other effects also being taken into account. Additionally, it has a wavelet implementation to deal with the red noise component in light curves, using the \citet{2009ApJ...704...51C} model. %It is also possible to fit radial velocity and photometric data simultaneously. %including an implementation of the Rossiter-McLaughlin effect (not finalized yet). 
The global minimum of the $\chi^2$ or $\log L$ values are searched for by a genetic algorithm, with a subsequent refinement of the fit by an annealing algorithm. Error estimation is handled by a (differential evolution) Markov chain Monte Carlo algorithm.

TLCM takes the light curves as an input and fits for the epoch of the transit, orbital period (if multiple transits are available), semi-major axis, planet-to-star radius ratio, impact parameter and the limb-darkening parameters. In this regard, we mainly fit for the orbital period and mid-transit time to use them as priors for our RV modelling. Input parameters for our system were taken from TEPCat \citep{2011MNRAS.417.2166S}.

We also tried searching the TESS photometry for evidence of rotational variability. Using Lomb-Scargle periodograms, we searched the whole TESS light curve, as well as each sector separately, after removing the transits, for periodic modulations. This search yielded no conclusive rotation period for WASP-106 (\citealt{2014AA...570A..64S} also found no rotational variability), and so we are unable to convert our sky-projected obliquity to a true obliquity measurement.

%------------------------------------------------------------------

\section{Results}\label{sec:results}

\begin{figure}[h]
    \centering
    \includegraphics[width=\linewidth]{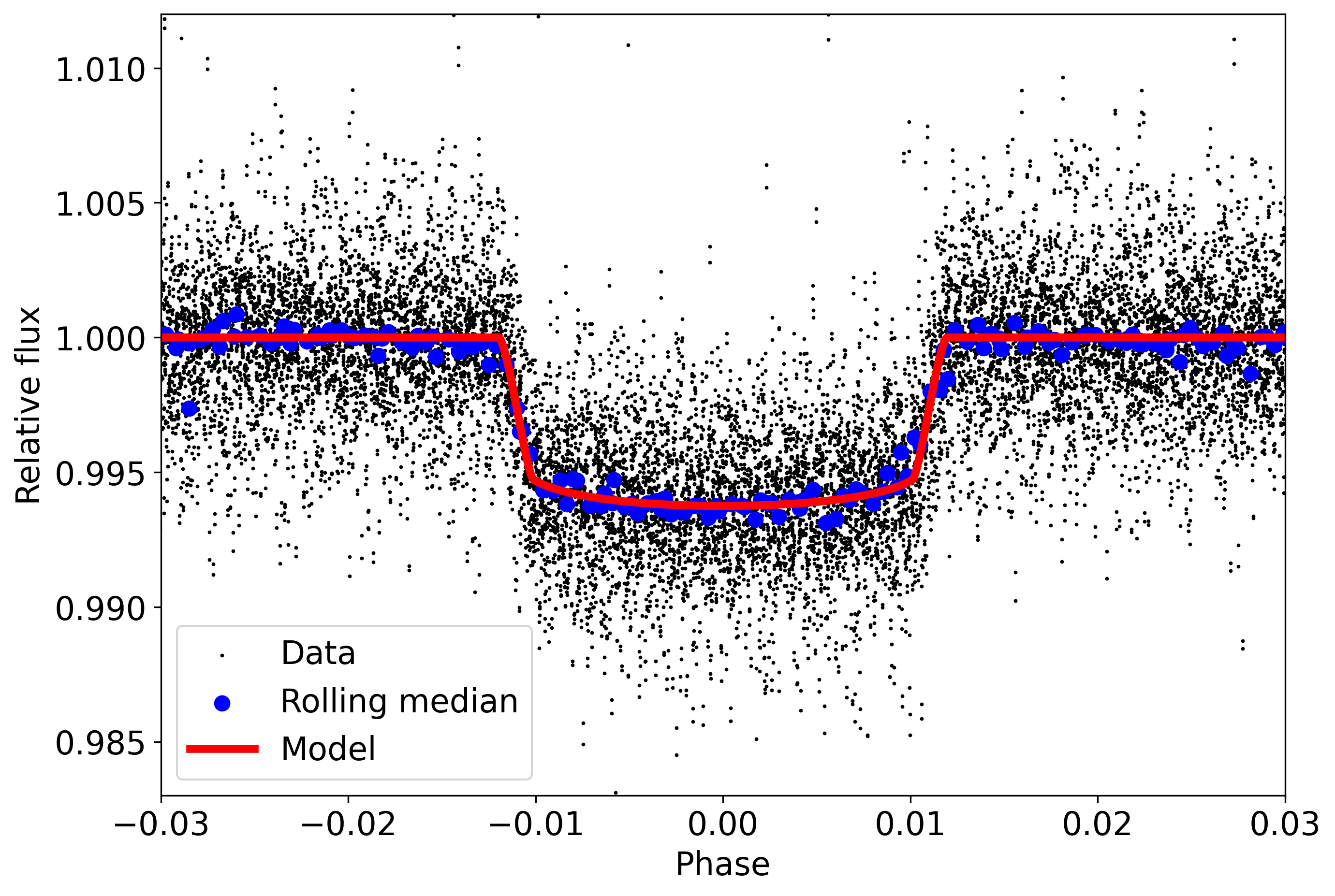}
    \caption{Phase-folded transit for all available TESS sectors. The TESS data is shown as black dots, the rolling median is shown as blue dots, and the final TLCM model is shown in red.}
    \label{fig:transit}
\end{figure}

Using TLCM to fit the TESS data, we obtain the mid-transit time and orbital period reported in Table~\ref{tab:priors_results} to use them as priors for the fit to the RV data. A plot of the phase-folded data and model are shown in Fig.~\ref{fig:transit}.

The results from the MCMC optimisation for the RV data are given in Tab.~\ref{tab:priors_results}.
From our RV fit, we find a small eccentricity with:
\begin{eqnarray*}
    \sqrt{e}\,\cos\omega &= -0.171 \pm 0.015\\
    \sqrt{e}\,\sin\omega &=  0.152 \pm 0.028.
\end{eqnarray*}
Solving this system of equations leads to $e = 0.05\pm0.01$. To test whether the elliptical or circular orbit is the preferred solution, we repeat the fit to the radial velocity data with the eccentricity fixed at zero. Comparing the Bayesian Information Criterion (BIC) values for the respective fits, we obtain $\mathrm{BIC_c} = 8632$ for the circular orbit and $\mathrm{BIC_e} = 8100$, indicating that the eccentric orbit is preferred. To further validate this result, we employ the F-test of \citet{1971AJ.....76..544L}, and find a probability of just $1.7\%$ for the apparent eccentricity to be spurious. According to their limit of $5\%$ and in combination with the other evidence, we conclude that there is only a very small chance that the measured eccentricity could arise from a truly circular orbit, and hence we adopt the eccentric solution.

Aside from this, and as our main result, we find the projected stellar obliquity to be $\lambda = (-1 \pm 11)^\circ$.
This means that the stellar spin axis is aligned with the planetary orbital axis. 
Furthermore, we improve the uncertainties on some of the previously measured parameters from the aforementioned studies of this system, including a more precise measurement of the RV semi-amplitude as well as an updated orbital period.

%\begin{table}[h]
%    \centering
%    \renewcommand{\arraystretch}{1.1} % Default value: 1
%    \caption{Results for our MCMC modelling of the RM data. $T_\mathrm{c}$ and $P$ are inferred from the photometric data. $T_\mathrm{c}$ is given in BJD$_\mathrm{TDB}$ - 2450000. The subscripts ``H1'', ``H2'', ``HN'', ``CO'' and ``SO'' stand for HARPS first observation, HARPS second observation, HARPS-N, CORALIE and SOPHIE, respectively.}
%    \begin{tabular}{l c c}
%        \hline
%        Parameter [unit] & Value & Error \\\hline
%        $K$ [m\,s$^{-1}$] & 162.4888 & 1.4907 \\
%        $\sqrt{e}\cos\omega$ & -0.1706 & 0.0148 \\
%        $\sqrt{e}\sin\omega$ & 0.1515 & 0.0286 \\
%        $v\sin i$ [m\,s$^{-1}$] & 6714.0813 & 251.7864 \\
%        $\lambda$ [$^\circ$] & -1.05 & 11.31 \\
%        $u_1$ & 0.7868 & 0.1089 \\
%        $u_2$ & 0.4004 & 0.1381 \\
%        $a\,R_\star^{-1}$ & 14.0411 & 0.2796 \\
%        $b$ & 0.0359 & 0.0917 \\
%        $R_\mathrm{p}\,R_\star^{-1}$ & 0.0780 & 0.0010 \\
%        $T_\mathrm{c}$ [d] & 9297.11710 & 0.00076 \\
%        $P$ [d] & 9.289718 & 0.000005 \\
%        $\gamma_\mathrm{H1}$ [m\,s$^{-1}$] & 17260.12 & 1.22 \\
%        $\gamma_\mathrm{H2}$ [m\,s$^{-1}$] & 17237.48 & 0.87 \\
%        $\gamma_\mathrm{HN}$ [m\,s$^{-1}$] & 17259.87 & 1.16 \\
%        $\gamma_\mathrm{CO}$ [m\,s$^{-1}$] & 17248.13 & 4.27 \\
%        $\gamma_\mathrm{SO}$ [m\,s$^{-1}$] & 17190.67 & 5.65 \\
%         \hline
%    \end{tabular}
%    \label{tab:results}
%\end{table}

\begin{figure*}[]
    \centering
    \includegraphics[width=0.8\textwidth]{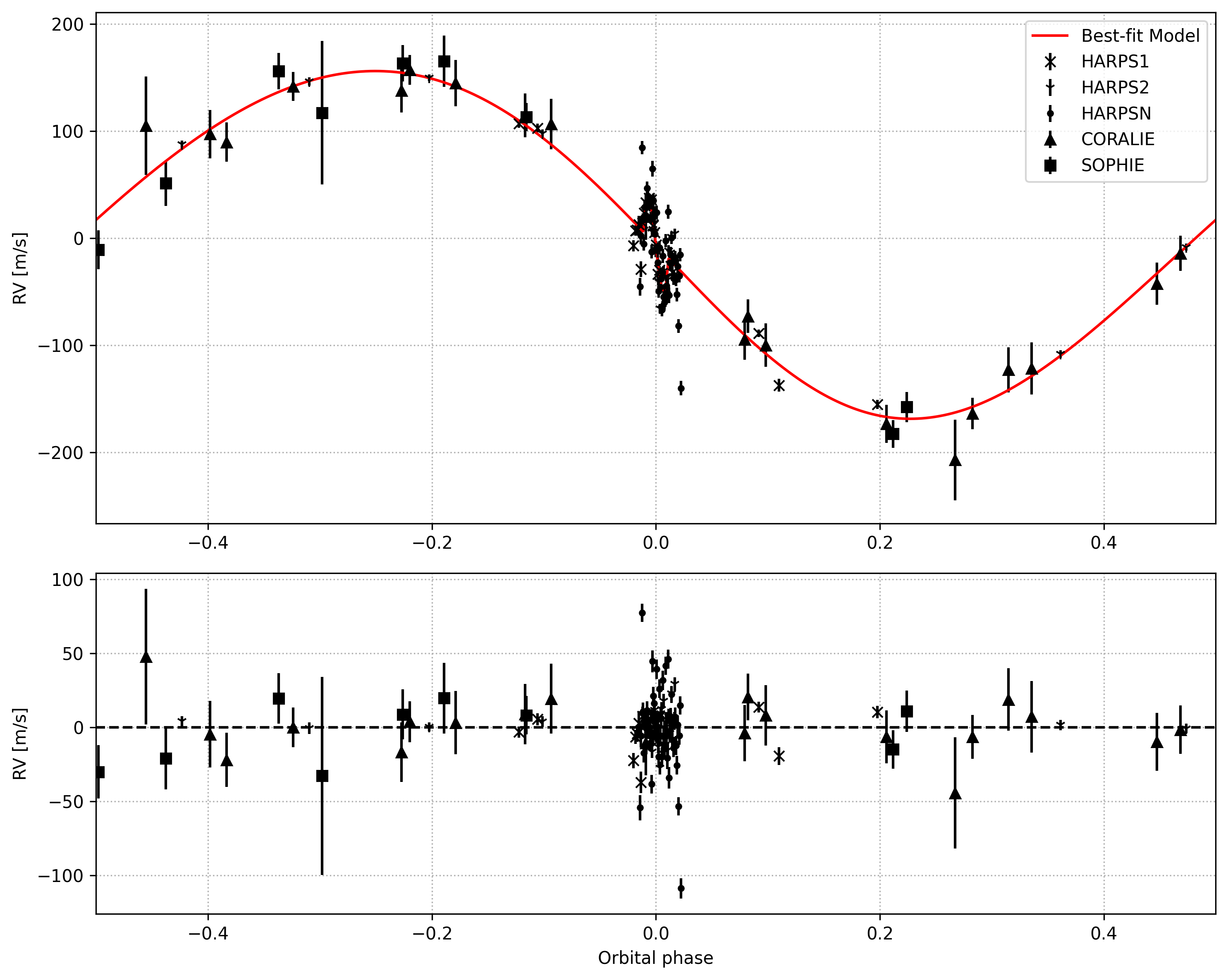}
    \caption{Top: RVs and best-fit model for all of our data sets of WASP-106, showing the Keplerian orbit. The orbital phase is shown on the x-axis, with the radial velocity being shown on the y-axis. The different instruments are marked according to the legend, with ``HARPS1'' and ``HARPS2'' describing the first and second observation made with the HARPS spectrograph. The best-fit model is shown as the red curve. Bottom: Residuals of our best-fitting model.}
    \label{fig:Keplerian}
\end{figure*}

\begin{figure*}[t]
    \centering
    \includegraphics[width=0.8\textwidth]{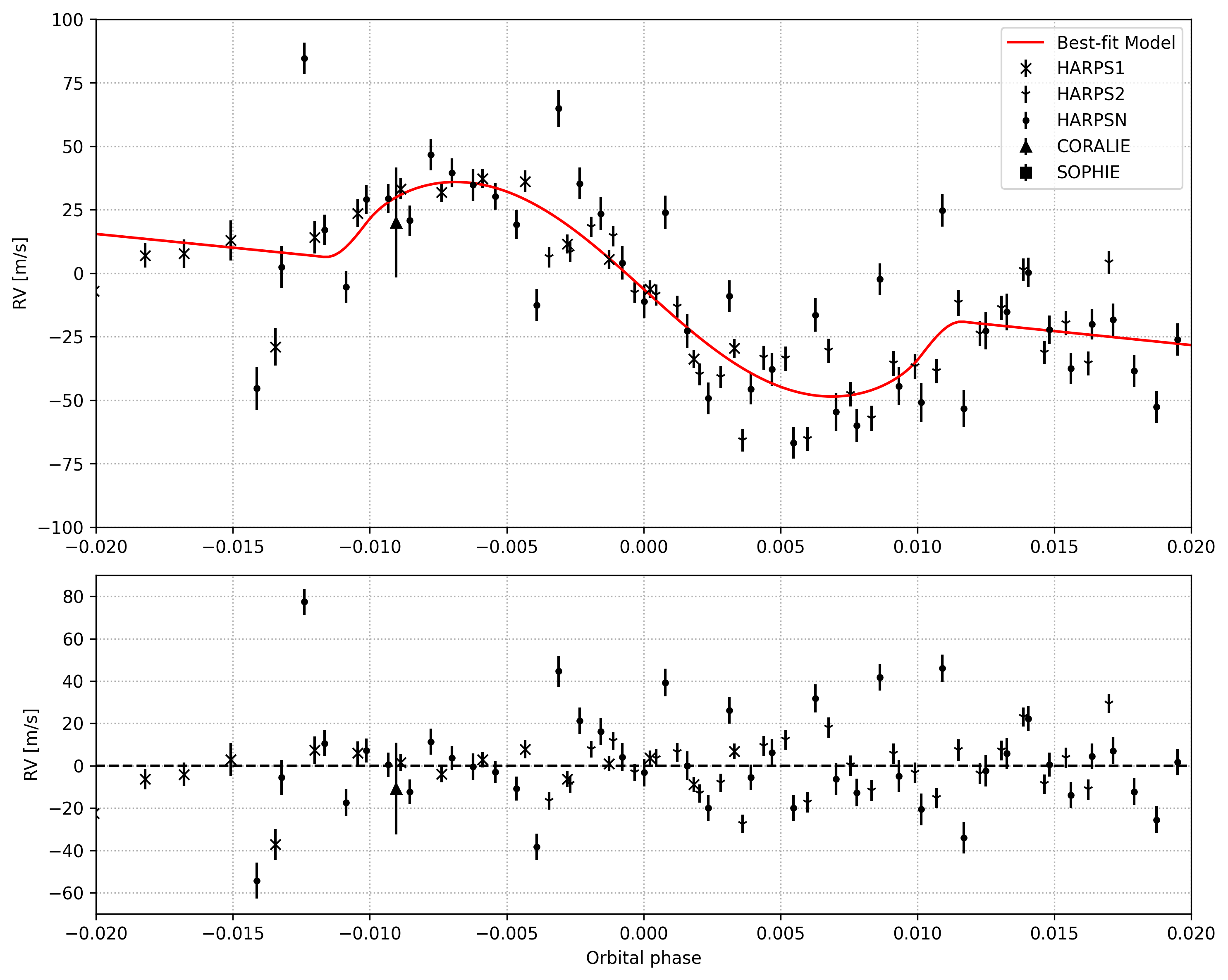}
    \caption{Same as Fig.~\ref{fig:Keplerian}, but zoomed-in on the Rossiter-McLaughlin anomaly.}
    \label{fig:RM}
\end{figure*}

The full Keplerian orbit in the RVs is shown in Fig.~\ref{fig:Keplerian} together with the residuals from the fit.
A zoom-in of the Rossiter-McLaughlin anomaly is shown in Fig.~\ref{fig:RM}. The effect is clearly visible.

%------------------------------------------------------------------

\section{Discussion}\label{sec:disc}

From our measurement of the stellar obliquity angle $\lambda$, we find the system to be aligned. This is in accordance with the trend that was observed in single star systems by \citet{2022AJ....164..104R}. The true 3D obliquity $\psi$ of the system could be different however, since we can only measure the sky projection of this angle using the technique in this paper. 
The relatively large error bar that we obtain from our measurement is mainly caused by the scatter in the RVs during transit. A possible cause of this could be clouds during the observations. 
%For the HARPS observations, we checked the ESO weather archive for the La Silla site, which indicates ... for the first HARPS observations and ... for the second HARPS observation, based on ... \JH{put footnote here with the link(s)}. In addition, we also checked a further website\JH{link}, which indicates ... for the specific nights. For the HARPS-N observations, we checked the weather stations of nearby telescopes \JH{link here}. These show some scatter in the seeing, which could indicate clouds. However, there is also a large (~ X hour) gap in the records for the night of the observations. Moreover, we also looked at the ``SKY BACKGROUND'' parameter during the observations in the FITS headers. These suggest .... \JH{Look at those} .
%For the radial velocity measurements taken during transit of the planet in front of the star, 
It seems like that for both of the nights in early 2015, there were passing clouds in the sky\footnote{based on the seeing (\href{https://www.not.iac.es/weather/dimm_index.php?day=16&month=01&year=2015&submit=Raw+Data}{Jan 2015}, \href{https://www.not.iac.es/weather/dimm_index.php?day=13&month=02&year=2015&submit=Raw+Data}{Feb 2015}) and according to the historical weather data website \href{https://www.timeanddate.com/weather/@2513466/historic?month=2\&year=2015}{timeanddate.com}.}, which could have impaired the observations and led to the greater scatter for the data sets labelled as ``HARPS2'' and ``HARPSN'' in Fig.~\ref{fig:RM}. This scatter hinders a more precise retrieval of the obliquity angle $\lambda$.
Removing both affected data sets, leaving only the first night of the HARPS RM observations (and a single CORALIE measurement), which covers approximately three quarters of the full transit, results in $\lambda = (-6.73\pm16.97)^\circ$. This result is consistent with the result from the fit to the full data set, and so is also consistent with alignment.
Upon searching the residual RVs for additional periodic trends after subtracting our best-fit model, we find no additional signals which could correspond to further planets in the system.

The relatively small measured eccentricity, which we validated by comparing BIC values of two fits with free and fixed eccentricities and also the F-test of \citet{1971AJ.....76..544L}, is an indication for the planetary orbit to be largely unaffected by tidal interactions between the planet and its host star, which could be an indication for our measured projected obliquity to be primordial. 
This supports the hypothesis of quiescent migration through the disk for warm Jupiters, in contrast to hot Jupiters, which are believed to get into their tight orbits via high-eccentricity migration due to interactions with other bodies in the early system (see e.g. \citealt{2022PASP..134h2001A, 2022ApJ...926L..17R, 2022AJ....164..104R}).
%Our updated system parameters \JH{are in agreement?} with previous studies, except for \JH{insert here if necessary, could also discriminate between 1,2,3 sigma etc.}
Even though we delivered the first measurement of the spin-orbit alignment in this system, it will likely not be the last, since recently, new observations were taken using ESPRESSO. These measurements should be able to constrain $\lambda$ even more. %, and might also allow a first measurement of the true 3D obliquity.
Aside from using the HARPS DRS RV data, we also extracted the RVs using SERVAL, which leads to a similar value of $\lambda$ with an equally-sized error bar.
%Furthermore, we also fitted the measured RVs in combination with the photometric data using TLCM as an additional comparison, even though the implementation of the Rossiter-McLaughlin effect in TLCM is still in development. The results are in agreement with each other, even though the error bar on $\lambda$ seems rather small, which is why we do not adopt this value. \JH{Should this be in the paper at all?}

We provide updated radial velocity and transit parameters, which are mostly consistent with those from the literature within $1\,\sigma$. In more detail, for our determined RV semi-amplitude ($K$), we find an uncertainty that is about a third compared to the value found in \citet{2014AA...570A..64S}. This improvement on the uncertainty is mainly driven by the few HARPS data points that extend beyond the transit phase of the planet.
%\JH{Put actual values to have a direct comparison or is this already enough?}. 
The same is the case for our measurement of the stellar rotational velocity ($v\sin i$). 
Since we do not fix the eccentricity in the RV fit, due to the availability of more measurements that have been taken in the meanwhile, we find a non-zero eccentricity at $e = 0.05$, in comparison to previous studies of this system. 
Moreover, our retrieved values for the scaled semi-major axis ($a\,R_\star^{-1}$) and impact parameter ($b$) from the RV data and the photometric data from TESS are in good agreement with each other and with those from \citet{2014AA...570A..64S}, but they only agree with the values of \citet{2021MNRAS.506.3810B} within $4\,\sigma$ and $6\,\sigma$, respectively. However, the high impact parameter from the latter study compensates for the lower semi-major axis value, caused by the well known degeneracy between the two parameters (see e.g. \citealt{2020A&A...640A.134A}).
% TLCM and RV agree well + Smith14
% Borsato high b compensates for the lower a/Rs value (well known degeneracy between the two parameters)
%A similar and even more pronounced case of this can be seen when comparing the impact parameters. Most likely, this is due to the different observing bands of the RV instruments and CHEOPS. 
Apart from the good agreement of the planet-to-star radius ratio ($R_\mathrm{P}R_\star^{-1}$), the transit center time ($T_\mathrm{c}$), and the orbital period ($P$), for the latter of which we now find a smaller uncertainty, there is a small difference between the systemic velocities measured here and in \citet{2014AA...570A..64S}. 
Mainly, we find greater error bars for the offsets of CORALIE and SOPHIE. Nevertheless, their RV offsets lie within $1\,\sigma$ of our determined offsets.
The values we find from our analysis of the TESS data with TLCM are in good agreement with those of \citet{2021MNRAS.506.3810B}. Our derived mid-transit time is within $2\,\sigma$ of their value, although they have a slightly smaller uncertainty because of the more precise CHEOPS data that they used in their analysis. The orbital periods agree well within $1\,\sigma$. However, we are able to further improve the uncertainty of the latter due to the many transits that are available in the TESS data.

%WASP-106\,b RM observations with HARPS-N: It was probably cloudy in the night, as is also a bit apparent in the data (from: https://www.timeanddate.com/weather/@2513466/historic?month=2\&year=2015) \JH{Is there an official weather source for TNG at Roque de los Muchachos?}. This leads to the relatively large scatter for the data taken in the second half of the night and hence the error bar on the measured obliquity angle. However, alignment of this system can still be confirmed and is supported by the general trend of aligned warm Jupiters in single star systems, as has been highlighted by \citet{2022AJ....164..104R}.\\
% \JH{Mention the small eccentricity -> e ~ 0.0574 (error?), $\omega$ ~ -0.7205 (error?), only "weak" interactions leave the "primordial" eccentricity more or less untouched}
%\JH{Mention the recent ESPRESSO observations of WASP-106!}

%------------------------------------------------------------------

\section{Conclusions\label{sec:conclusion}}

From our analysis of three separate radial velocity data sets during transit of WASP-106\,b in front of its host star, two of which were taken with HARPS, and one of which was taken with HARPS-N, we provide the first measurement of the stellar obliquity in this system, with $\lambda = (-1\pm11)^\circ$, indicating good alignment between the stellar spin axis and the planetary orbital axis. 
More precise radial velocity measurements during transit of the planet might be able to put even tighter constraints on the orbital obliquity in this system, since the two HARPS RM observations only cover the first and the second half of the observed transit, respectively. The measurements from the second observation taken with HARPS, and those from HARPS-N show in addition to this scatter in the second half of the transit, which limits the precision with which the orbital obliquity can be inferred. This scatter could be caused by the presence of clouds, as the weather data suggest. 

Our work adds another aligned system to the sample of 16 warm Jupiter systems with measured spin-orbit angles. Of these, 14 are aligned and two are misaligned at about $-30^\circ$ (HAT-P-17, \citealt{2022A&A...664A.162M}) and $40^\circ$ (TOI-1859, \citealt{2023arXiv230516495D}).
Some of these planets, even though they show alignment, have non-zero eccentricities. These can usually be attributed to undetected companion bodies in the system. The same could be the case for our system with the small, but non-zero eccentricity for the planet's orbit that we find at high confidence, which is validated by the BIC values of respective fits, and also the \citet{1971AJ.....76..544L} F-test for small eccentricities.
However, by examining the Lomb-Scargle periodogram \citep{1976Ap&SS..39..447L, 1982ApJ...263..835S}, we do not find signs of other bodies. Given the moderate orbital period of about 9.3\,d, the eccentricity damping time scale may be relatively long for WASP-106\,b, making it possible for the eccentricity to be of primordial origin.
% How many are aligned warm Jupiters in single-star systems?
This supports the theory that warm Jupiters migrate quiescently through the disk, in contrast to hot Jupiters which are thought to get into their tight orbits via high-eccentricity migration. 
%A recent paper by \citet{2023arXiv230516495D} that highlights a misaligned warm Jupiter on a highly eccentric orbit, disfavours the general trend normally found in these systems.

%In addition to this, we also measure a small, but non-zero eccentricity for the planet's orbit at high confidence. This is validated by the BIC values of respective fits, and also the \citet{1971AJ.....76..544L} F-test for small eccentricities.

%------------------------------------------------------------------

\begin{acknowledgements}
%We thank the anonymous referee for their helpful comments.

JVH acknowledges the support of the DFG priority programme SPP 1992 “Exploring the Diversity of Extrasolar Planets (SM 486/2-1)”.
This work was supported by European Union’s Horizon Europe Framework Programme under the Marie Skłodowska-Curie Actions  grant agreement No. 101086149 (EXOWORLD).
The results in this work are partly based on observations made with ESO Telescopes at the La Silla Paranal Observatory under programme IDs 093.C-0474(A) and 094.C-0090(A).
This work is based on observations made with the Italian Telescopio Nazionale Galileo (TNG) operated on the island of La Palma by the Fundación Galileo Galilei of the INAF (Istituto Nazionale di Astrofisica) at the Spanish Observatorio del Roque de los Muchachos of the Instituto de Astrofísica de Canarias under programme ID OPT14B\_66.
%\JH{The results in this work are partly based on observations made with the INAF TNG telescope under programme ID OPT14B\_66.}
This paper includes data collected with the TESS mission, obtained from the MAST data archive at the Space Telescope Science Institute (STScI). Funding for the TESS mission is provided by the NASA Explorer Program. STScI is operated by the Association of Universities for Research in Astronomy, Inc., under NASA contract NAS 5–26555.
%This work made use of TEPCat to collect system properties \citep{2011MNRAS.417.2166S}.
This research received funding from the European Research Council (ERC) under the European Union's Horizon 2020 research and innovation programme (grant agreement n$^\circ$ 803193/BEBOP). This work made use of Astropy:\footnote{http://www.astropy.org} a community-developed core Python package and an ecosystem of tools and resources for astronomy \citep{2013A&A...558A..33A, 2018AJ....156..123A, 2022ApJ...935..167A}.

\end{acknowledgements}

\bibliographystyle{aasjournal}
\bibliography{literature}

\appendix
\section{Radial velocity data}

\begin{table}[h]
    \setlength{\tabcolsep}{12pt}
    \centering
    \caption{Radial velocity data of WASP-106\,b. The 'Observation' column refers from 1-5 to the data sets from SOPHIE, CORALIE, HARPS (first night), HARPS (second night), and HARPS-N, respectively. Time is given in BJD$_\mathrm{TDB} - 2450000$.}
    \begin{tabular}{c c c c}
        \hline
        Time [d]   & RV [km$\,$s$^{-1}$]  & Error [km$\,$s$^{-1}$] & Observation \\\hline
        6298.61990 & 17.033000 & 0.014000 & 1           \\
        6329.63330 & 17.242000 & 0.021000 & 1           \\
        6330.56960 & 17.347000 & 0.017000 & 1           \\
        6331.59700 & 17.354000 & 0.017000 & 1           \\
        6360.49150 & 17.304000 & 0.013000 & 1           \\
        6363.53450 & 17.008000 & 0.013000 & 1           \\
        6364.68432 & 17.126600 & 0.024250 & 2           \\
        6365.72574 & 17.205720 & 0.019670 & 2           \\
        6366.62637 & 17.353310 & 0.045860 & 2           \\
        6368.74628 & 17.385940 & 0.020170 & 2           \\
        6369.77261 & 17.362930 & 0.020360 & 2           \\
        6370.77349 & 17.268240 & 0.021700 & 2           \\
        6371.59523 & 17.153570 & 0.018930 & 2           \\
        6371.76896 & 17.148240 & 0.020370 & 2           \\
        6372.76894 & 17.074720 & 0.017800 & 2           \\
        6373.78397 & 17.125310 & 0.021030 & 2           \\
        6377.37630 & 17.308000 & 0.067000 & 1           \\
        6378.38890 & 17.356000 & 0.024000 & 1           \\
        6403.39260 & 17.180000 & 0.018000 & 1           \\
        6410.50052 & 17.041100 & 0.037690 & 2           \\
        6423.58725 & 17.390130 & 0.013520 & 2           \\
        6424.55331 & 17.405420 & 0.013750 & 2           \\
        6438.51331 & 17.084330 & 0.014740 & 2           \\
        6441.47794 & 17.345440 & 0.022550 & 2           \\
        6443.51263 & 17.392980 & 0.021490 & 2           \\
        6449.52837 & 17.234020 & 0.016300 & 2           \\
        6469.48415 & 17.337830 & 0.018350 & 2           \\
        6481.46628 & 17.354740 & 0.023680 & 2           \\
        6696.76093 & 17.175350 & 0.015760 & 2           \\
        6750.59786 & 17.366900 & 0.003779 & 3           \\
        6750.75405 & 17.362900 & 0.004046 & 3           \\
        6751.54932 & 17.253000 & 0.005156 & 3           \\
        6751.56677 & 17.267100 & 0.004778 & 3           \\
        6751.57991 & 17.267800 & 0.005576 & 3           \\
        6751.59571 & 17.273000 & 0.007835 & 3           \\
        6751.61093 & 17.231100 & 0.007376 & 3           \\
        6751.62421 & 17.274200 & 0.006412 & 3           \\
        6751.63890 & 17.283700 & 0.005411 & 3           \\
        6751.65343 & 17.293300 & 0.004182 & 3           \\
        6751.66727 & 17.291900 & 0.003870 & 3           \\\hline
    \end{tabular}
    \label{tab:RVs}
\end{table}

\setcounter{table}{3}
\begin{table*}[h]
    \centering
    \setlength{\tabcolsep}{12pt}
    \caption{\textit{(continued)}}
    \begin{tabular}{c c c c}
        \hline
        Time [d]   & RV [km$\,$s$^{-1}$]  & Error [km$\,$s$^{-1}$] & Observation \\\hline
        6751.68125 & 17.297400 & 0.003607 & 3           \\
        6751.69565 & 17.296200 & 0.004319 & 3           \\
        6751.70992 & 17.271600 & 0.003740 & 3           \\
        6751.72418 & 17.265500 & 0.003590 & 3           \\
        6751.73801 & 17.253800 & 0.003509 & 3           \\
        6751.75282 & 17.226300 & 0.003559 & 3           \\
        6751.76665 & 17.230500 & 0.003609 & 3           \\
        6752.59058 & 17.171300 & 0.003583 & 3           \\
        6752.75786 & 17.122600 & 0.006005 & 3           \\
        6753.57398 & 17.104900 & 0.004204 & 3           \\
        7033.78505 & 17.129400 & 0.003421 & 4           \\
        7034.82784 & 17.229000 & 0.003252 & 4           \\
        7035.78606 & 17.325100 & 0.003365 & 4           \\
        7036.84025 & 17.383800 & 0.003938 & 4           \\
        7037.83560 & 17.387300 & 0.003011 & 4           \\
        7038.77477 & 17.335300 & 0.003641 & 4           \\
        7039.68490 & 17.243700 & 0.004058 & 4           \\
        7039.69209 & 17.245800 & 0.004005 & 4           \\
        7039.69934 & 17.255700 & 0.003966 & 4           \\
        7039.70678 & 17.252100 & 0.004099 & 4           \\
        7039.71402 & 17.229800 & 0.003994 & 4           \\
        7039.72127 & 17.228900 & 0.004167 & 4           \\
        7039.72852 & 17.224300 & 0.004361 & 4           \\
        7039.73604 & 17.197600 & 0.004329 & 4           \\
        7039.74316 & 17.196600 & 0.004313 & 4           \\
        7039.75061 & 17.171600 & 0.004430 & 4           \\
        7039.75779 & 17.204200 & 0.004678 & 4           \\
        7039.76526 & 17.203800 & 0.004747 & 4           \\
        7039.77263 & 17.172200 & 0.004725 & 4           \\
        7039.77982 & 17.206900 & 0.004758 & 4           \\
        7039.78726 & 17.189800 & 0.004773 & 4           \\
        7039.79444 & 17.180300 & 0.004962 & 4           \\
        7039.80183 & 17.202000 & 0.004882 & 4           \\
        7039.80914 & 17.200800 & 0.004851 & 4           \\
        7039.81638 & 17.198900 & 0.004816 & 4           \\
        7039.82385 & 17.225800 & 0.005114 & 4           \\
        7039.83109 & 17.213700 & 0.004898 & 4           \\
        7039.83848 & 17.223800 & 0.004747 & 4           \\
        7039.84581 & 17.238800 & 0.004507 & 4           \\
        7039.85297 & 17.206200 & 0.004647 & 4           \\
        7039.86030 & 17.217800 & 0.004858 & 4           \\
        7039.86782 & 17.201900 & 0.004709 & 4           \\
        7039.87485 & 17.241600 & 0.004556 & 4           \\
        7067.45494 & 17.214500 & 0.008478 & 5           \\
        7067.46341 & 17.262300 & 0.008211 & 5           \\
        7067.47106 & 17.344400 & 0.006175 & 5           \\
        \hline
    \end{tabular}
    \label{tab:RVs}
\end{table*}

\setcounter{table}{3}
\begin{table*}[h]
    \centering
    \setlength{\tabcolsep}{12pt}
    \caption{\textit{(continued)}}
    \begin{tabular}{c c c c}
        \hline
        Time [d]   & RV [km$\,$s$^{-1}$]  & Error [km$\,$s$^{-1}$] & Observation \\\hline
        7067.47799 & 17.276900 & 0.006085 & 5           \\
        7067.48528 & 17.254500 & 0.006337 & 5           \\
        7067.49218 & 17.288900 & 0.005686 & 5           \\
        7067.49953 & 17.289300 & 0.005745 & 5           \\
        7067.50681 & 17.280600 & 0.005943 & 5           \\
        7067.51400 & 17.306500 & 0.006212 & 5           \\
        7067.52121 & 17.299400 & 0.005694 & 5           \\
        7067.52830 & 17.294600 & 0.006239 & 5           \\
        7067.53587 & 17.290100 & 0.005165 & 5           \\
        7067.54307 & 17.279000 & 0.005667 & 5           \\
        7067.54993 & 17.247300 & 0.006303 & 5           \\
        7067.55735 & 17.324800 & 0.007338 & 5           \\
        7067.56451 & 17.295200 & 0.006260 & 5           \\
        7067.57171 & 17.283300 & 0.006428 & 5           \\
        7067.57894 & 17.263900 & 0.006614 & 5           \\
        7067.58634 & 17.248800 & 0.006575 & 5           \\
        7067.59357 & 17.283800 & 0.006579 & 5           \\
        7067.60091 & 17.237200 & 0.006699 & 5           \\
        7067.60812 & 17.210600 & 0.006265 & 5           \\
        7067.61531 & 17.250800 & 0.006215 & 5           \\
        7067.62263 & 17.214200 & 0.006029 & 5           \\
        7067.62974 & 17.222000 & 0.006404 & 5           \\
        7067.63704 & 17.193100 & 0.006250 & 5           \\
        7067.64444 & 17.243400 & 0.006597 & 5           \\
        7067.65143 & 17.205200 & 0.007483 & 5           \\
        7067.65846 & 17.199900 & 0.006507 & 5           \\
        7067.66630 & 17.257500 & 0.006203 & 5           \\
        7067.67276 & 17.215400 & 0.007515 & 5           \\
        7067.68041 & 17.209000 & 0.007598 & 5           \\
        7067.68754 & 17.284600 & 0.006480 & 5           \\
        7067.69477 & 17.206500 & 0.007343 & 5           \\
        7067.70217 & 17.237200 & 0.007430 & 5           \\
        7067.70942 & 17.244600 & 0.007264 & 5           \\
        7067.71668 & 17.260200 & 0.005846 & 5           \\
        7067.72383 & 17.237600 & 0.005626 & 5           \\
        7067.73110 & 17.222400 & 0.006139 & 5           \\
        7067.73823 & 17.239800 & 0.006056 & 5           \\
        7067.74544 & 17.241600 & 0.006365 & 5           \\
        7067.75258 & 17.221400 & 0.006386 & 5           \\
        7067.76016 & 17.207200 & 0.006372 & 5           \\
        7067.76726 & 17.233700 & 0.006322 & 5           \\
        7067.77435 & 17.177800 & 0.006230 & 5           \\
        7067.78161 & 17.224800 & 0.006180 & 5           \\
        7067.78878 & 17.244300 & 0.006133 & 5           \\
        7067.79602 & 17.119800 & 0.006872 & 5          \\\hline
    \end{tabular}
    \label{tab:RVs}
\end{table*}

\end{document}